\documentclass[twocolumn,showpacs,aps,prl,superscriptaddress,floatfix]{revtex4}

\usepackage{graphicx}
\usepackage{dcolumn}
\usepackage{amsmath}
\usepackage{epsfig}

\RequirePackage{xspace}

\usepackage{relsize}
\def\babar{\mbox{\slshape B\kern-0.1em{\smaller A}\kern-0.1em
    B\kern-0.1em{\smaller A\kern-0.2em R}}}

\def\epem       {\ensuremath{e^+e^-}\xspace}

\def\Kbar  {\kern 0.2em\overline{\kern -0.2em K}{}\xspace}

\def\Kz    {\ensuremath{K^0}\xspace}
\def\Kzb   {\ensuremath{\Kbar^0}\xspace}
\def\KzKzb {\ensuremath{\Kz \kern -0.16em \Kzb}\xspace}
\def\Kp    {\ensuremath{K^+}\xspace}
\def\Km    {\ensuremath{K^-}\xspace}

\def\KpKm  {\ensuremath{\Kp \kern -0.16em \Km}\xspace}
 
\def\KL    {\ensuremath{K^0_{\scriptscriptstyle L}}\xspace}

\def\Dbar    {\kern 0.2em\overline{\kern -0.2em D}{}\xspace}

\def\Dz      {\ensuremath{D^0}\xspace}
\def\Dzb     {\ensuremath{\Dbar^0}\xspace}
\def\DzDzb   {\ensuremath{\Dz {\kern -0.16em \Dzb}}\xspace}
\def\Dp      {\ensuremath{D^+}\xspace}
\def\Dm      {\ensuremath{D^-}\xspace}

\def\DpDm    {\ensuremath{\Dp {\kern -0.16em \Dm}}\xspace}

\def\Bbar    {\kern 0.18em\overline{\kern -0.18em B}{}\xspace}

\def\BB      {\ensuremath{B\Bbar}\xspace} 
\def\Bz      {\ensuremath{B^0}\xspace}
\def\Bzb     {\ensuremath{\Bbar^0}\xspace}
\def\BzBzb   {\ensuremath{\Bz {\kern -0.16em \Bzb}}\xspace}
\def\Bu      {\ensuremath{B^+}\xspace}
\def\Bub     {\ensuremath{B^-}\xspace}

\def\BpBm    {\ensuremath{\Bu {\kern -0.16em \Bub}}\xspace}

\def\BorBbar    {\kern 0.18em\optbar{\kern -0.18em B}{}\xspace}
\def\DorDbar    {\kern 0.18em\optbar{\kern -0.18em D}{}\xspace}
\def\KorKbar    {\kern 0.18em\optbar{\kern -0.18em K}{}\xspace}

\mathchardef\Upsilon="7107
\def\Y#1S{\ensuremath{\Upsilon{(#1S)}}\xspace}

\def\FourS {\Y4S}

\mathchardef\Deltares="7101
\mathchardef\Xi="7104
\mathchardef\Lambda="7103
\mathchardef\Sigma="7106
\mathchardef\Omega="710A

\def\Deltabar{\kern 0.25em\overline{\kern -0.25em \Deltares}{}\xspace}
\def\Lbar{\kern 0.2em\overline{\kern -0.2em\Lambda\kern 0.05em}\kern-0.05em{}\xspace}
\def\Sigbar{\kern 0.2em\overline{\kern -0.2em \Sigma}{}\xspace}
\def\Xibar{\kern 0.2em\overline{\kern -0.2em \Xi}{}\xspace}
\def\Obar{\kern 0.2em\overline{\kern -0.2em \Omega}{}\xspace}
\def\Nbar{\kern 0.2em\overline{\kern -0.2em N}{}\xspace}
\def\Xb{\kern 0.2em\overline{\kern -0.2em X}{}\xspace}

\def\mes        {\mbox{$m_{\rm ES}$}\xspace}

\newcommand{\tev}{\ensuremath{\mathrm{\,Te\kern -0.1em V}}\xspace}
\newcommand{\gev}{\ensuremath{\mathrm{\,Ge\kern -0.1em V}}\xspace}
\newcommand{\mev}{\ensuremath{\mathrm{\,Me\kern -0.1em V}}\xspace}
\newcommand{\kev}{\ensuremath{\mathrm{\,ke\kern -0.1em V}}\xspace}
\newcommand{\ev}{\ensuremath{\mathrm{\,e\kern -0.1em V}}\xspace}
\newcommand{\gevc}{\ensuremath{{\mathrm{\,Ge\kern -0.1em V\!/}c}}\xspace}
\newcommand{\mevc}{\ensuremath{{\mathrm{\,Me\kern -0.1em V\!/}c}}\xspace}
\newcommand{\gevcc}{\ensuremath{{\mathrm{\,Ge\kern -0.1em V\!/}c^2}}\xspace}
\newcommand{\mevcc}{\ensuremath{{\mathrm{\,Me\kern -0.1em V\!/}c^2}}\xspace}

\def\invfb   {\ensuremath{\mbox{\,fb}^{-1}}\xspace}

\def\mus  {\ensuremath{\rm \,\mus}\xspace}

\def\mus        {\ensuremath{\,\mu{\rm s}}\xspace}    

\def\ra                 {\ensuremath{\rightarrow}\xspace}
\def\to                 {\ensuremath{\rightarrow}\xspace}

\def\pep2{PEP-II}

\def\gsim{{~\raise.15em\hbox{$>$}\kern-.85em
          \lower.35em\hbox{$\sim$}~}\xspace}
\def\lsim{{~\raise.15em\hbox{$<$}\kern-.85em
          \lower.35em\hbox{$\sim$}~}\xspace}

\xspace

\newcommand{\nimBaseC}       {Nucl.\ Instr.\ and Methods\xspace}

\newcommand{\nim}       [1]  {\nimBaseC~{\bf #1}}

\def\jetset74   {\mbox{\tt Jetset \hspace{-0.5em}7.\hspace{-0.2em}4}\xspace}

\newcommand{\BABARPubYear}    {04}
\newcommand{\BABARPubNumber}  {032}

\newcommand{\SLACPubNumber} {10662}

\def\eeul      {\ensuremath {6.1 \times10^{-8}}}
\def\mmul      {\ensuremath {8.3 \times10^{-8}}}
\def\emul      {\ensuremath {18 \times10^{-8}}}
\def\nexpee    {\ensuremath {0.71 \pm 0.31}}
\def\nexpmm    {\ensuremath {0.72 \pm 0.26}}
\def\nexpem    {\ensuremath {1.29 \pm 0.44}}

\def\ldata     {\ensuremath {111 \invfb}}
\def\bll {\ensuremath {\Bz \to \ell^{+} \ell^{-}}}
\def\bee {\ensuremath {\Bz \to e^{+} e^{-}}}
\def\bmm {\ensuremath {\Bz \to \mu^{+} \mu^{-}}}
\def\bem {\ensuremath {\Bz \to e^{\pm} \mu^{\mp}}}

\def\ie {{\it i.e.}}
\def\etal {{\it et al.}}
\def\de        {\ensuremath {\Delta E}}

\def\figurebox#1#2#3{%
    \def\arg{#3}%
    \ifx\arg\empty
    {\hfill\vbox{\hsize#2\hrule\hbox to #2{\vrule\hfill\vbox to #1{\hsize#2\vfill}\vrule}\hrule}\hfill}%
    \else
    {\hfill\epsfbox{#3}\hfill}%
    \fi}

\begin{document}


\begin{flushleft}
\babar-PUB-\BABARPubYear/\BABARPubNumber\\
SLAC-PUB-\SLACPubNumber\\
\end{flushleft}

{\normalsize 

\title{{ Search for decays of {\boldmath $B^0$} mesons into pairs of charged 
leptons: \\
{\boldmath $B^0 \ra e^+e^-$, $B^0 \ra \mu^+\mu^-$, $B^0 \ra e^{\pm}\mu^{\mp}$} }}

%
\author{B.~Aubert}
\author{R.~Barate}
\author{D.~Boutigny}
\author{F.~Couderc}
\author{J.-M.~Gaillard}
\author{A.~Hicheur}
\author{Y.~Karyotakis}
\author{J.~P.~Lees}
\author{V.~Tisserand}
\author{A.~Zghiche}
\affiliation{Laboratoire de Physique des Particules, F-74941 Annecy-le-Vieux, France }
% ...
\collaboration{The \babar\ Collaboration}
\noaffiliation

\author{\large \babar\ Collaboration}

\date{\today}

\begin{abstract}
We present a  search for the decays $B^0 \rightarrow
e^+ e^-$, $B^0 \rightarrow \mu^+\mu^-$, and $\bem$
in data collected  at the $\FourS$ resonance with the  \babar\ detector at the
SLAC $B$ Factory. Using a data set  of \ldata, we find no evidence for 
a signal in any of the three channels investigated and set the following 
branching fraction upper 
limits at the  $90\%$ confidence level:  ${\cal B}(\bee)< \eeul$, 
${\cal B}(\bmm) < \mmul$, and ${\cal B}(\bem) < \emul$.
\end{abstract}


\maketitle

In the Standard Model (SM), rare $B$ decays such as $\bll$, where 
$\ell$ refers to $e$ or $\mu$,  are expected to proceed through 
diagrams such as those shown in Fig.~\ref{fig:bllfeyn} 
(charge conjugate processes are included implicitly throughout).
\begin{figure}[t]
\begin{center}
\vskip 0.2cm
\includegraphics[width=5.3cm]{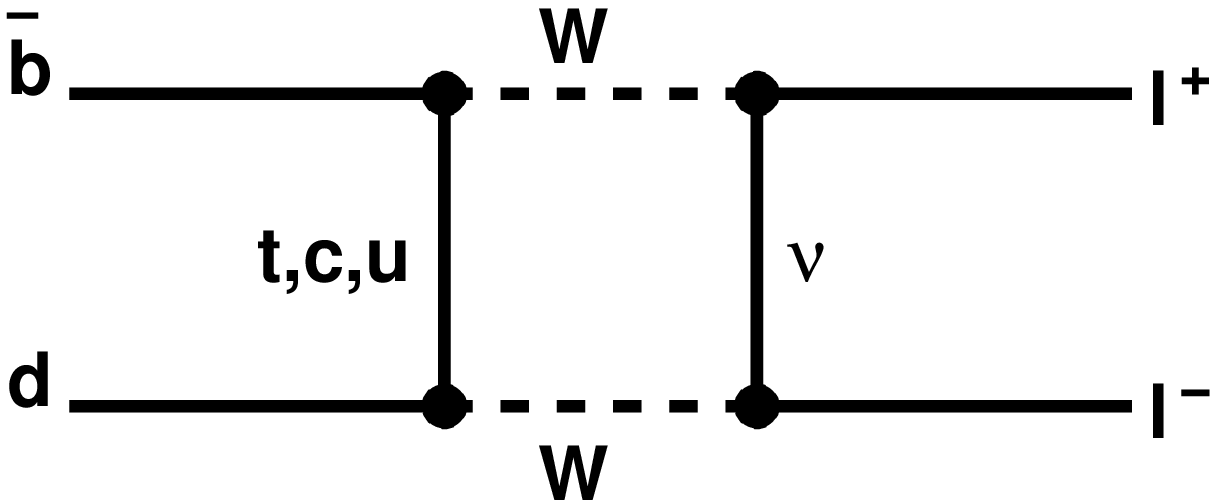}
\vskip 0.2cm
\includegraphics[width=5.3cm]{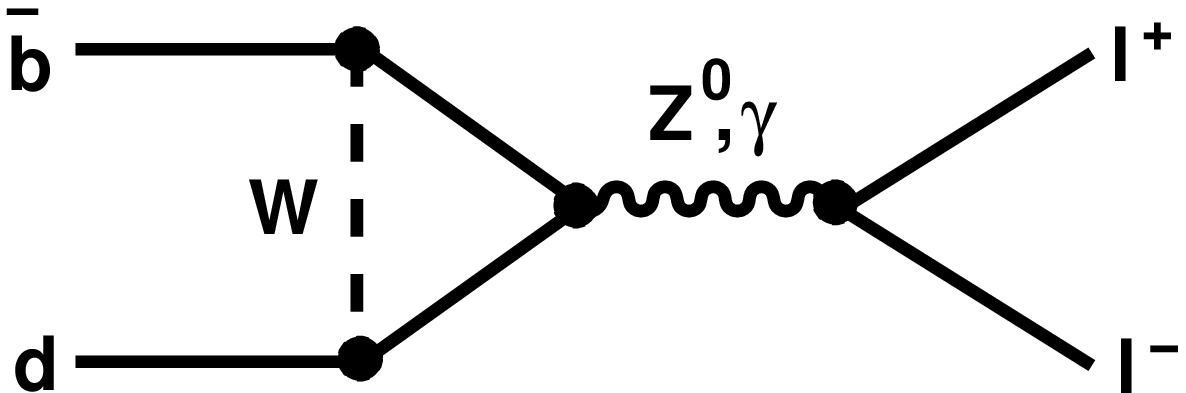}
\end{center}
\caption {Representative Feynman diagrams for $\bll$ in the Standard Model. } 
\label{fig:bllfeyn}
\end{figure}
These decays are highly suppressed since they involve a $b \to d$  
transition and require an internal quark annihilation within the $B$ 
meson.
In addition, the decays are
helicity suppressed by factors of $(m_{\ell}/m_B)^{2}$ where $m_\ell$ is the
mass of the lepton $\ell$ and $m_B$ is the mass of the $B$ meson. 
$B^0$ decays to leptons 
of two different flavors, violate lepton flavor conservation and are
therefore forbidden in the SM, although permitted in extensions to 
the SM with non-zero neutrino mass~\cite{numass}. 
The SM expectations are given in Table~\ref{tab:expectedBR}.

Since these processes are highly suppressed in the SM, they are potentially 
sensitive probes of physics beyond the SM. 
In the minimally supersymmetric Standard
Model (MSSM) the branching fraction for these decays can be enhanced by
orders of magnitude~\cite{SUSY}. In particular, for MSSM models with
modified minimal flavor violation ${\rm (\overline{MFV})}$ and large 
tan$\beta$~\cite{bobeth},
the branching fraction can be increased by up to four orders of magnitude.
Experimental bounds can restrict allowed regions of parameter space, 
specifically the mass of the charged Higgs boson.
In non-MSSM models 
with two Higgs doublets and natural flavor
conservation at large $\tan \beta$, an increase in the branching fraction of
several orders of magnitude is expected~\cite{KOLDA}. 
$B^0 \to \ell^+ \ell^-$ decays are also
allowed in specific models containing leptoquarks~\cite{Davidson:1993qk} and
supersymmetric (SUSY) models without R-parity~\cite{ROY}. 
The branching fractions for the flavor violating channels 
$B^0 \to \ell^+_i \ell^-_j$ ($i \not = j$) are expected to be exceedingly 
small but can be enhanced by leptoquarks or
$R$-parity violating operators in SUSY models.

To date, \bll\ decays have not been observed.
As  shown in Table~\ref{tab:expectedBR},  experimental limits are 
approaching a level of sensitivity that will restrict
the allowed parameter space of models that
produce \bll \ branching fraction enhancements of a few orders of magnitude 
with respect to the SM rates.

\begin{table}[t]
\begin{center}
\caption{The expected branching fractions in the 
Standard Model~\cite{SMTheory} and the current best upper limits (U.L.) at the
$90\%$  C.L.}
\vspace{0.1in}
\begin{small}
\begin{tabular}{llccc} \hline 
Decay & \multicolumn{1}{c}{SM} & CLEO~\cite{cleo00} & Belle~\cite{belle03} 
      & CDF~\cite{cdf04} 
\\ 
Mode & Expectation      &$9.1\invfb$
                        &$78\invfb$
                        &$0.17\invfb$
\\\hline
&&&&\\[-0.3cm]
$e^+e^-$ &$1.9\times10^{-15}$ & $8.3 \times10^{-7}$
         &$1.9\times10^{-7}$  
         & -- \\ 
$\mu^+\mu^-$ &$8.0\times10^{-11}$ & \ $6.1 \times10^{-7}$ \  
         & \ $1.6 \times10^{-7}$ \ 
         & \ $1.5 \times10^{-7}$ \  \\
$e^{\pm}\mu^{\mp}$ &\multicolumn{1}{c}{--} & $15 \times10^{-7}$ 
         &$1.7 \times10^{-7}$ 
         & --  \\ 
\hline
\end{tabular}
\end{small}
\label{tab:expectedBR}
\end{center}
\end{table}

The data used in these analyses were collected with the \babar\ detector
at the \pep2\ $\epem$ storage ring and
correspond to an integrated luminosity of $111~\invfb$ 
accumulated at the \FourS\ resonance (``on-resonance'') and 
$11.9~\invfb$ accumulated at a 
center-of-mass (c.m.) energy about $40\mev$ below the \FourS\ resonance
(``off-resonance''). The latter sample is used for non-resonant $q\bar q$
($q = u, d, s,$ and $c$) background studies.
The collider is operated with asymmetric beam energies, 
producing a boost ($\beta\gamma = 0.55$) of the \FourS\ along the 
collision axis.  

The \babar\ detector is optimized for the asymmetric beam
configuration at PEP-II and is described in detail in \cite{babarnim}.
The 1.5-T superconducting solenoidal magnet, whose cylindrical volume is 
1.4~m in radius and 3~m long, contains a charged-particle 
tracking system, a Cherenkov detector dedicated to 
charged-particle identification, and central and forward electromagnetic
CsI calorimeters (EMC). 
The segmented flux return, including endcaps, is instrumented with 
resistive plate chambers for muon and \KL identification.

The presence of two charged high-momentum leptons provides 
a very clean signature for the three decay modes under consideration.
We require two oppositely-charged 
high-momentum leptons (\ie \ $|p^*_\ell| \sim m_B/2$ where $p^*_\ell$ is the 
c.m. momentum of lepton $\ell$) from a common vertex 
consistent with the decay of a $B^0$ meson.
Since the signal events contain two $B^0$ mesons and no additional particles,
the total energy of each $B^0$ in
the c.m. must be equal to half of the total beam energy.
We define
\begin{eqnarray}
m_{\rm ES}&=& \sqrt{(E_{\rm beam}^*)^2 - (\sum_i {\bf p}^*_i)^2} 
\label{eq1} \\
\Delta E&=& \sum_i\sqrt{m_i^2 + ({\bf p}_i^*)^2}- E_{\rm beam}^*,
\end{eqnarray}
where $E_{\rm beam}^*$ is the ($e^+$ or $e^-$) beam energy in the c.m. frame,
${\bf p}_i^*$ is the momentum of lepton $i$ in the
c.m. frame, and $m_i$ is the mass of lepton $i$. 
In Eq.~(\ref{eq1}), $E_{\rm beam}^*$ is used as opposed to $E_B^*$ because
$E_{\rm beam}^*$ is known with much greater precision.
For correctly reconstructed $B^0$ mesons,
$m_{\rm ES}$ peaks at the mass of the $B^0$ meson with a resolution of about 
2.5 \mevcc and $\Delta E$ peaks near zero.

To reduce background from lepton misidentification, we require the leptons
to satisfy stringent
electron and muon identification criteria~\cite{incl_charm}. 
The electron identification efficiency is greater than $93\%$ with a
mis-identification rate of less than $0.3\%$.
The muon identification efficiency ranges
from $(55 - 70)\%$ (depending on run period) with a mis-identification rate
of 3\%.
Electron energy lost through bremsstrahlung is partially recovered by 
adding the energy of photons that lie within a 3 degree cone about the 
electron direction.

Suppression of background from non-resonant $q\bar q$
production is provided by a series of topological requirements.
In particular,  we require $|\cos\theta_T| <  0.8$, where $\theta_T$
is the angle in the c.m. frame between the thrust axis of the particles 
that form the
reconstructed $\Bz$  candidate and the thrust axis  of the remaining
tracks and neutral clusters in the event.
In addition we employ cuts on the invariant mass of the ``Rest Of the Event'' 
(all tracks not associated with the $\Bz$ candidate where all non-leptonic
tracks are assumed to be pions) of
$m_{\rm ROE} > 0.5$~GeV and on the second normalized 
Fox-Wolfram moment of $R_2 < 0.8$~\cite{foxwolf}.
We also cut on the total multiplicity of both charged tracks and neutral
particles by means of the variable $N_{\rm mult}$ defined as 
$N_{\rm mult} = N_{\rm trk} + N_\gamma/2,$
where $N_{\rm trk}$ is the
total number of tracks in the event and $N_\gamma$ is the number of photons
found with an energy $E_\gamma > 80$ MeV. We require $N_{\rm mult} \geq 5.5$
for the $ee$ and $e\mu$ channels and $N_{\rm mult} \geq 5.0$ for the $\mu\mu$
channel. This variable is especially useful
in the rejection of radiative Bhabha events. We also require that the total
energy in the EMC ($E_{\rm EMC}$) be less than 
11 GeV. This cut is effective in reducing background from QED $e^+e^-$
events, including radiative Bhabhas with many conversions.

Four of the selection criteria given above 
($|\cos\theta_T|$, $m_{\rm ROE}$, $N_{\rm mult}$, and $E_{\rm EMC}$)
were simultaneously optimized for the best 
upper limit on ${\cal B}(\bll)$ where the assumed number of observed 
events is determined from
a Poisson distribution with the mean equal to the expected background.
Sideband data are compared with signal Monte Carlo (MC) for the $e^+e^-$
channel for 
four of these variables, $|\cos\theta_T|$, $m_{\rm ROE}$, $R_2$, and 
$N_{\rm mult}$ in Fig.~\ref{fig:vars}.

\begin{figure}
\centering
\includegraphics[width=4.2cm]{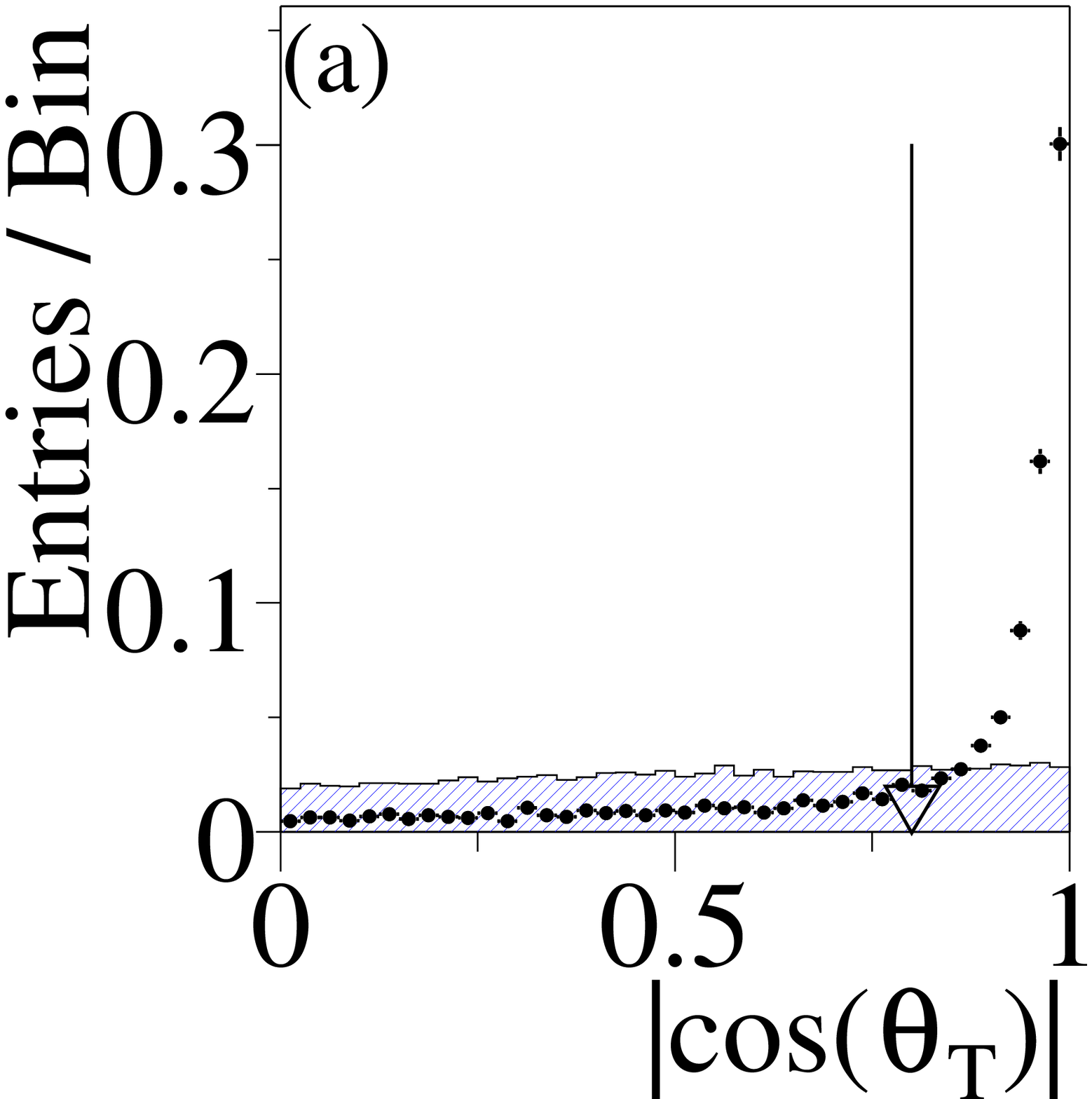}
\includegraphics[width=4.2cm]{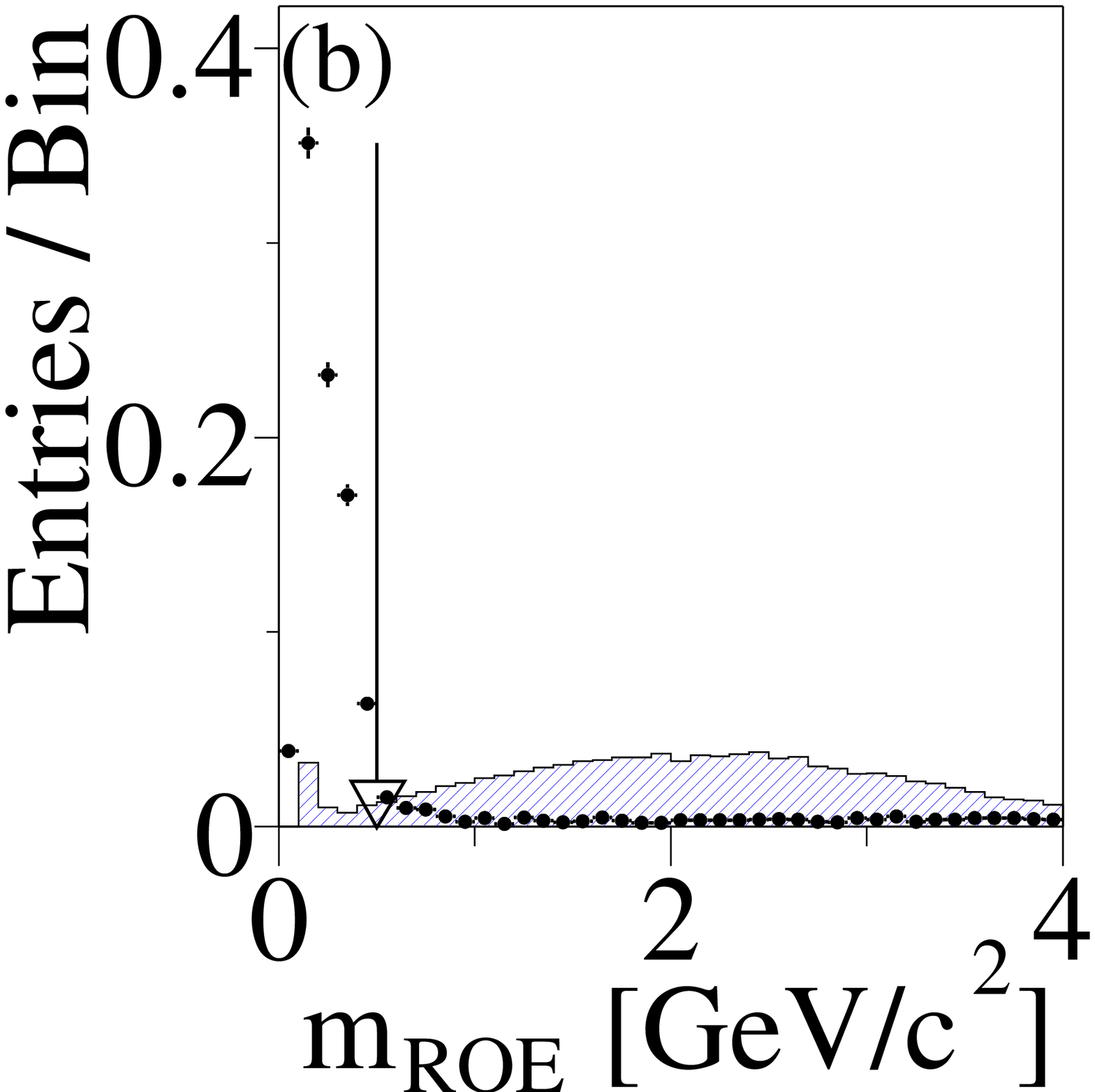}
\includegraphics[width=4.2cm]{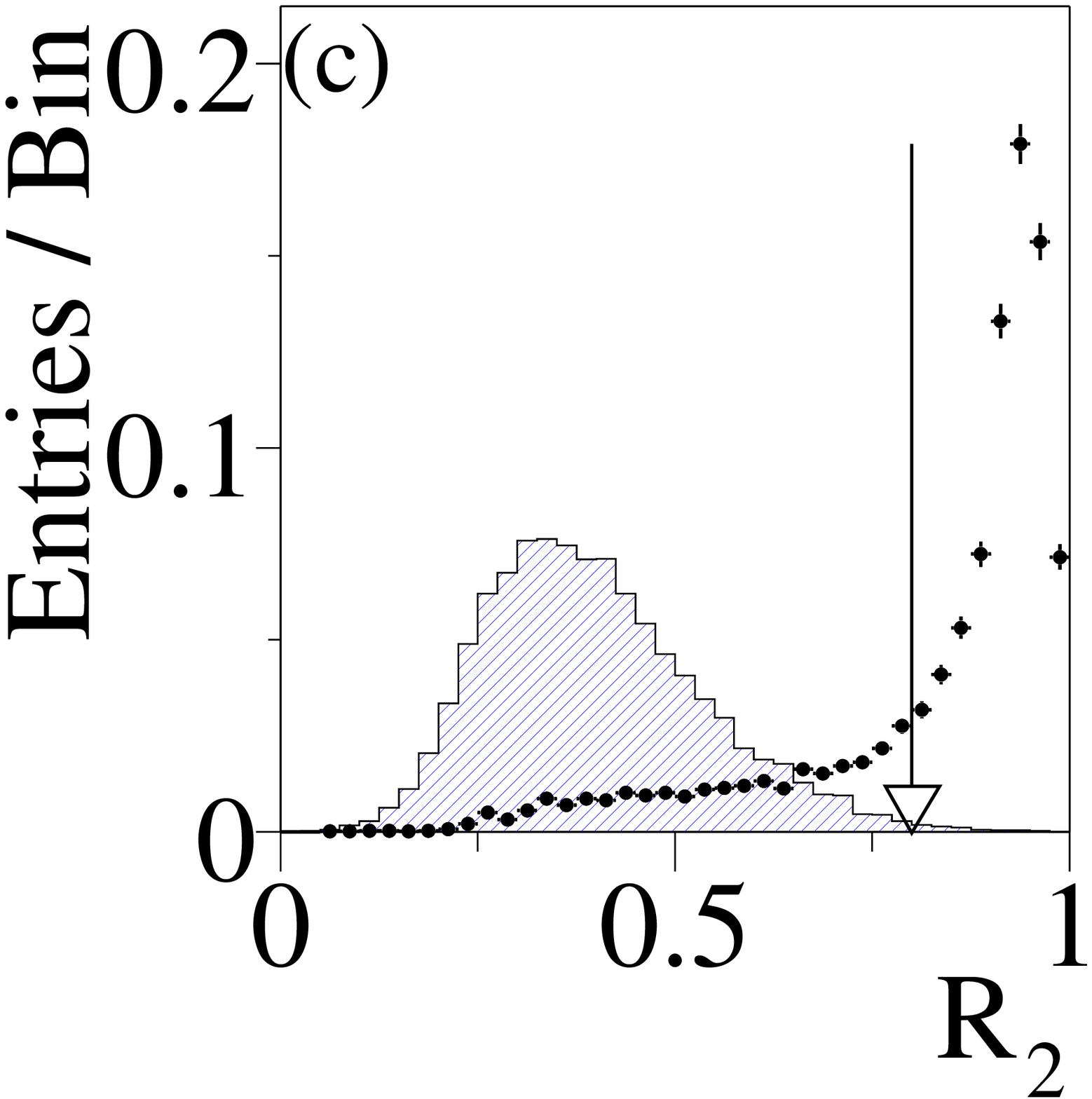}
\includegraphics[width=4.2cm]{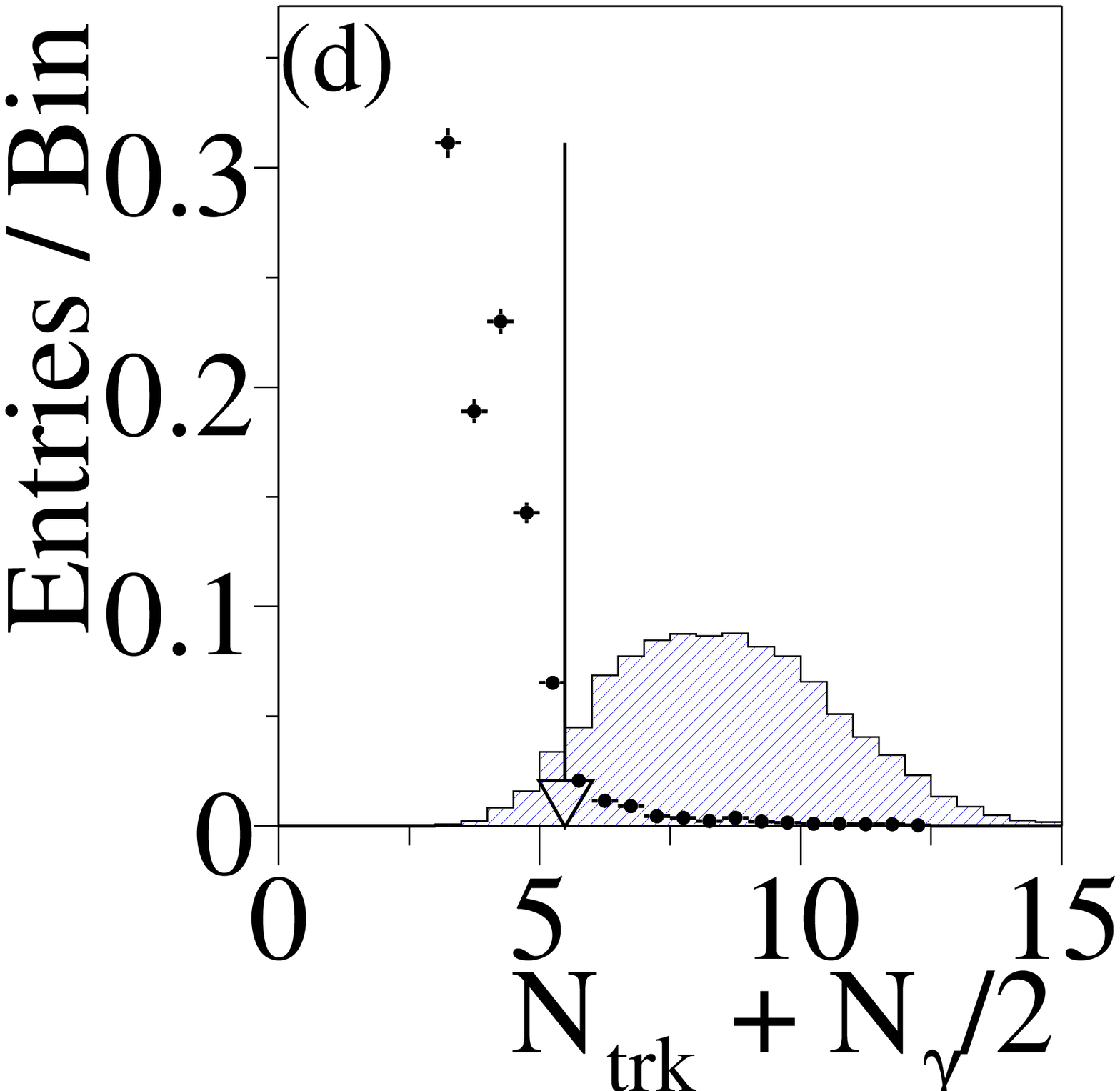}
\caption {Distributions of signal MC (hatched) and sideband data (points)
for the $e^+e^-$ channel after the initial selection cuts 
for (a) $|\cos\theta_T|$, 
(b) $m_{\rm ROE}$, (c) $R_2$, and (d) $N_{\rm mult}$. Arrows indicate 
final cut values. All distributions are normalized to unity. }
\label{fig:vars}
\end{figure}

\begin{table}[b]
\begin{center}
\caption{ Definition of the three different sideband boxes used for all
three decay modes.}
\vspace{0.1in}
\begin{small}
\begin{tabular}{lcc} \hline
Sideband Box & \ span in $\Delta E$ [GeV] \ 
& span in $m_{\rm ES}$ [GeV/$c^2$] \\ \hline
Grand Sideband     & $(-0.40, 0.40)$     & $(5.20, 5.26)$  \\
Upper $m_{\rm ES}$ & $ (0.20, 0.50)$ & $(5.20, 5.29)$  \\
Lower $m_{\rm ES}$ & $(-0.50, -0.20)$ & $(5.20, 5.29)$  \\
\hline
\end{tabular}
\end{small}
\label{tab:sb}
\end{center}
\end{table}

The $\bll$  candidates are selected by simultaneous
requirements on the energy difference $\de$ and the energy-substituted
mass $\mes$. For the $\bmm$ decay mode, the  size of this ``signal box'' 
is  chosen to be 
$[+2,-2]\sigma$ of the $\mes$ distribution and 
$[+2,  -2]\sigma$ for the $\de$.
In  the cases of the $\bee$  and $\bem$ decay modes, the
signal box sizes in $\mes$ are also $[+2,-2]\sigma$ but in
$\de$ are relaxed to $[+2, -3]\sigma$ and
$[+2,  -2.5]\sigma$, respectively, to accommodate the tail in the distribution 
resulting from uncorrected bremsstrahlung and final state radiation.
The resolution in $\mes$ is obtained from a fit to a Gaussian distribution, 
whereas the 
resolution in $\Delta E$ is obtained from a fit to an empirical 
function~\cite{novosibirsk} that gives a good description of this tail.

We estimate the background level in the signal box from the data sidebands
assuming that it is described by the ARGUS function~\cite{argus2} in $\mes$ 
and an exponential function in $\de$. We use these parameterizations to 
extrapolate the background level found in the sidebands into the signal box.
As indicated in Table~\ref{tab:sb}, three different sideband boxes are used.
The grand sideband box is used to estimate the functional form of the 
$\Delta E$ distribution. The upper and lower $m_{\rm ES}$ sideband boxes 
are used to estimate the functional form of the $m_{\rm ES}$ distribution.
Peaking backgrounds from misidentified two-body $B$ decay modes were 
estimated using an
MC sample equivalent to more than $20$ times the data luminosity and found 
to be negligible.
The total background expectations and signal efficiencies
are given in Table~\ref{tab:summary}.

The systematic uncertainties on the signal efficiency $\varepsilon$, 
the number of $B^0$ mesons produced in the data, and the background estimate
are incorporated into 
the determination of the upper limit on ${\cal B}(\bll)$. 
Since the signal efficiency is determined from
MC simulation only, differences between data and the simulation would result
in an error in our
normalization. To estimate this uncertainty we perform 
comparisons of data and MC using high statistics
control samples that have similar 
characteristics to our $\bll$ signal.
The optimal control samples are
$\Bz\to   J/\psi    K_S^0$,    with   $J/\psi\to    e^+e^-$ for $\bee$
and  $J/\psi\to    \mu^+\mu^-$ for $\bmm$, respectively.
Since there exists no appropriate control sample for the $e^\pm \mu^\mp$
mode, we use the larger of the systematic errors derived from either the
$ee$ or $\mu\mu$ modes.
In performing these comparisons we found a substantial uncertainty on the 
signal 
efficiency to be due to differences between data and the MC simulation in 
the mean and resolutions of various quantities, depending on the channel.
For the electron channels the dominant quantities are \de \ and $m_{\rm ROE}$
whereas for the muon channels they are $|\cos \theta_T|$, $N_{\rm mult}$, and
$m_{\rm ROE}$. When combined with the uncertainties on tracking efficiency
of 2.6\% and that for particle identification (1.0\% per electron, 3.0\% per
muon), the total systematic 
uncertainty on the efficiency is estimated to be 5.7\%, 7.1\%, and 6.8\%
for the $ee$, $\mu\mu$, and $e\mu$ modes respectively.

\begin{table}[t]
\begin{center}
\caption{Summary of the analyses where $N_{\rm obs}$ and $N_{\rm exp}^{\rm bg}$
are the observed and expected number of events in the signal box,
$\varepsilon$ is the efficiency, and  
${\cal B}_{\rm UL}(B^0\to \ell^+\ell^-)$ is the  upper limit on the branching 
fraction at the $90\%$ C.L. Systematic uncertainties on $N_{\rm exp}^{\rm bg}$ 
and $\varepsilon$ are given.}
\vspace{0.1in}
\begin{small}
\begin{tabular}{ccccc} \hline
Decay Mode&$N_{\rm obs}$   &$N_{\rm exp}^{\rm bg}$ &$\varepsilon [\%]$ &${\cal B}_{\rm UL}(B^0\to \ell^+\ell^-)
$ \\ \hline
&&&&\\[-0.3cm]
$e^+e^-$ & $0$ & \nexpee & $21.8 \pm 1.2 $ & \eeul \\
$\mu^+\mu^-$ & $0$ & \nexpmm & $15.9 \pm 1.1 $ & \mmul \\
$e^\pm\mu^\mp$ & $2$ & \nexpem & $18.1 \pm 1.2 $ & \emul \\
\hline
\end{tabular}
\end{small}
\label{tab:summary}
\end{center}
\end{table}

The background estimate is obtained from a fit to sideband data, so the 
primary uncertainty is due to fluctuations in the fit procedure
as events fall in or out of the sideband box. We  have  studied  the
stability  of the  fit  and  the background  estimate  when adding  or
removing events from  the \mes\ and \de\ histograms. 
We find that the fit is unbiased and stable to a level 
significantly less than the statistical uncertainty on the background 
estimate.

As shown in Fig.~\ref{fig:unblind} and Table~\ref{tab:summary}, 
when the contents of the signal box were revealed, 0, 0, and 2 events were 
found in the $ee$, $\mu\mu$, and $e\mu$ channels respectively.
As can be seen in Table~\ref{tab:summary}, the numbers of events 
found in the signal boxes are  compatible with the expected background
for each mode.

\begin{figure}[b]
\centering
\includegraphics[width=8.0cm]{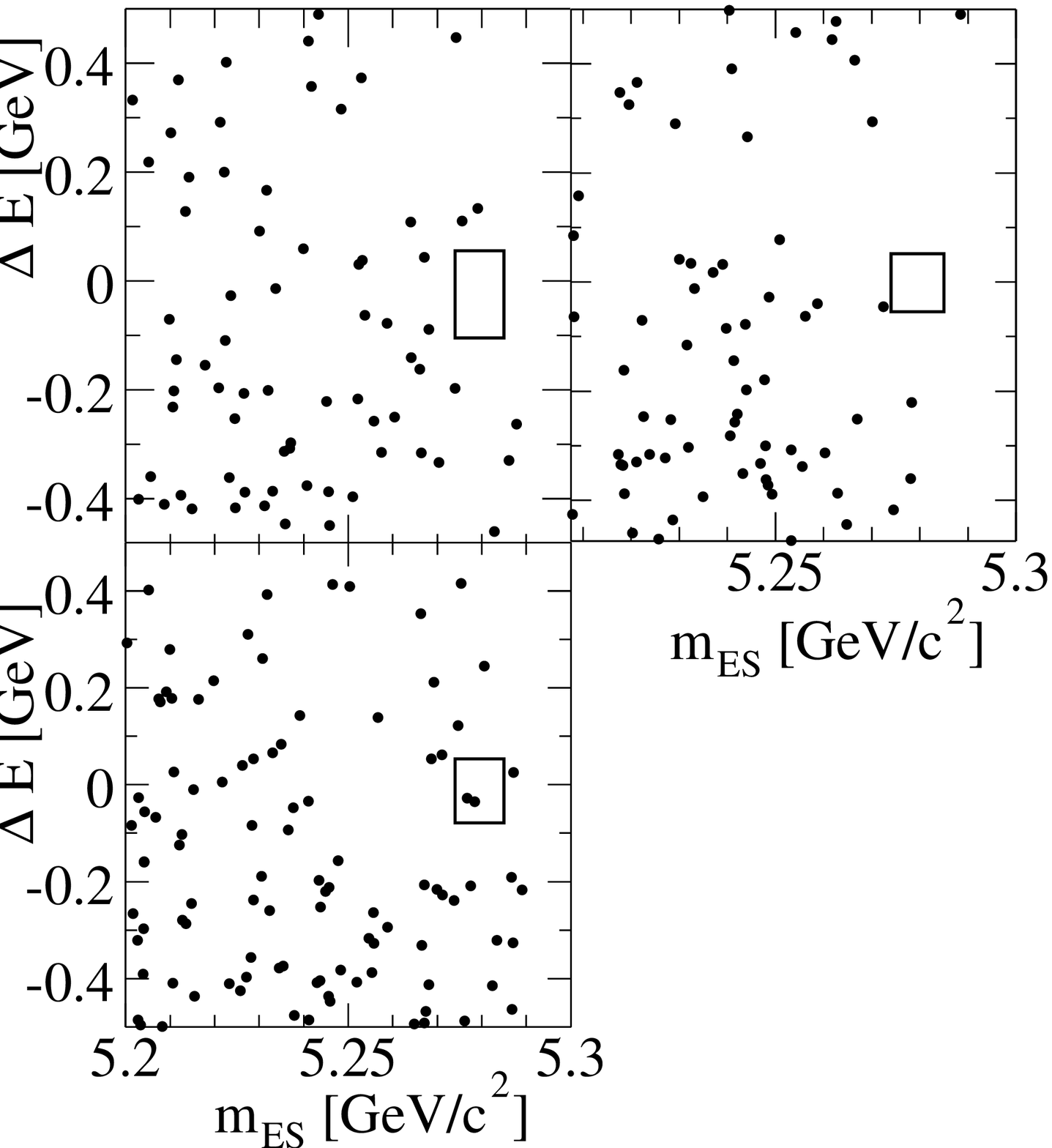}
\caption {Distribution of events in $\mes$ and $\de$ for $\bee$ (top left), 
$\bmm$ (top right), and $\bem$ (bottom).}
\label{fig:unblind}
\end{figure}

An upper limit on the branching fraction is computed using
\begin{eqnarray}
{\cal B}(\bll)={N_{UL}(N_{\rm obs})\over (N_{\Bz}+N_{\Bzb}) \cdot\varepsilon},
\end{eqnarray}
\noindent where  $N_{UL}(N_{\rm obs})$ is the Poisson 90\% U.L. on the number of
events assuming $N_{\rm obs}$ events have been observed,
$N_{\Bz}(N_{\Bzb})$ is the  number of $\Bz(\Bzb)$ mesons  produced in
the data, and  $\varepsilon$ is the signal  efficiency. We have
$N_{\Bz}  +  N_{\Bzb}  =   N_{\BB}$  under  the  assumption  of  equal
production  of  $\BzBzb$ and  $B^+B^-$  in  $\FourS$  decays. For our data
set $N_{\BB} = (122.5 \pm 1.2)\times 10^{6}$.

We follow the technique of \cite{barlow} 
in order to account for the presence of background and to 
include our systematic uncertainties in the determination of
the upper limit.
As summarized in Table~\ref{tab:summary}, the resulting upper 
limits at the 90\% C.L. for ${\cal B}(\bee)$, ${\cal B}(\bmm)$, and 
${\cal B}(\bem)$ are \eeul, \mmul, and \emul \ respectively.

These bounds are stringent enough to place interesting constraints on popular
models. For example, 
for the MSSM ${\rm (\overline{MFV})}$ models, the relation between ${\cal B}(\bmm)$
and the mass of the charged Higgs boson $m_H$ is given as a function of 
tan$\beta$ in~\cite{bobeth}. We find that for tan$\beta=60$, 
$m_H > 138$~GeV (90\% C.L.).

We are grateful for the excellent luminosity and machine conditions
provided by our \pep2\ colleagues, 
and for the substantial dedicated effort from
the computing organizations that support \babar.
The collaborating institutions wish to thank 
SLAC for its support and kind hospitality. 
This work is supported by
DOE
and NSF (USA),
NSERC (Canada),
IHEP (China),
CEA and
CNRS-IN2P3
(France),
BMBF and DFG
(Germany),
INFN (Italy),
FOM (The Netherlands),
NFR (Norway),
MIST (Russia), and
PPARC (United Kingdom). 
Individuals have received support from CONACyT (Mexico), A.~P.~Sloan Foundation, 
Research Corporation,
and Alexander von Humboldt Foundation.


\begin{thebibliography}{99}


\bibitem{numass} M. Maki, M. Nakagawa, and S. Sakata, 
Prog. Theor. Phys. {\bf 28}, 870 (1962); V.N. Gribov and B. Pontecorvo, 
Phys. Lett. {\bf B28}, 493 (1969). 


\bibitem{SUSY}
A. Dedes, H. K. Dreiner and U. Nierste,
Phys. Rev Lett. {\bf 87} (2001) 251804.


\bibitem{bobeth}
C.~Bobeth \etal, Phys. Rev. D {\bf 66}, 074021 (2002),
J. Urban, Proceedings of the 10th International Conference 
on Supersymmetry and Unification of Fundamental Interactions (SUSY02),
hep-ph/0210286.


\bibitem{KOLDA}
G. L. Kane, C. Kolda and J. E. Lennon,
hep-ph/0310042.


\bibitem{Davidson:1993qk}
S.~Davidson, D.~C.~Bailey and B.~A.~Campbell,
Z.\ Phys.\ C {\bf 61}, 613 (1994).

\bibitem{ROY}
D. Roy, Phys. Lett. {\bf B283} (1992) 270.


\bibitem{SMTheory}A. Ali, C. Greub, and T. Mannel, in {\it Proceedings
of the ECFA Workshop on the Physics of the European B Meson Factory},
edited by R. Aleksan and A. Ali, p. 155 (ECFA-93-151,C93/03/26), DESY-93-016.


\bibitem{cleo00} CLEO Collaboration, T. Bergfeld \etal, Phys. Rev. D{\bf62},
 {091102(R)} (2000).

\bibitem{belle03} Belle Collaboration, M.-C. Chang \etal, Phys. Rev. D{\bf68},
 {111101(R)} (2003).

\bibitem{cdf04} CDF Collaboration, D. Acosta \etal, hep-ex/0403032. 


\bibitem{babarnim}
\babar\ Collaboration, B.\ Aubert {\em et al.}, 
\nim {\bf A479}, 1 (2002).


\bibitem{incl_charm}
\babar\ Collaboration, B.\ Aubert {\em et al.}, 
Phys.\ Rev.\ D {\bf 67}, 032002 (2003). 


\bibitem{foxwolf} G.C. Fox and S. Wolfram, \prl {\bf 41} (1978) 1581. 


\bibitem{novosibirsk} \babar\ defines this empirical function to be \\
$f(E) = A\cdot \exp\left[-{1\over2}
\left({\log(1+\tau(E-\nu)\cdot{\sinh(\tau\sqrt{\log4})\over \sigma\tau\sqrt{\log4}})\over\tau}\right)^2+\tau^2\right],$ \\ 
where $\tau$ is the ``tail parameter'' 
(describing how  much is contained in the tail), $\sigma$ is the width, and 
$\nu$ is the peak position.


\bibitem{argus2}  ARGUS Collaboration, H.~Albrecht {\it et al.}, 
Phys. Lett. B {\bf 241}, 278 (1990); 

\bibitem{barlow} R.~Barlow,
Comput.\ Phys.\ Commun.\  {\bf 149}, 97 (2002). 


\end{thebibliography}
\end{document}